# Recent Decade's Power Outage Data Reveals the Increasing Vulnerability of U.S. Power Infrastructure

Bo Li[1*], Junwei Ma[2], Femi Omitaomu[3], Ali Mostafavi[4]

[1] Ph. D. student. Urban Resilience.AI Lab, Zachry Department of Civil and Environmental Engineering, Texas A&M University, College Station, Texas, United States.

[2] Ph. D. student. Urban Resilience.AI Lab, Zachry Department of Civil and Environmental Engineering, Texas A&M University, College Station, Texas, United States.

[3] Distinguished R&D Staff. Oak Ridge National Laboratory, Oak Ridge, Tennessee, United States.

[4] Professor. Urban Resilience.AI Lab, Zachry Department of Civil and Environmental Engineering, Texas A&M University, College Station, Texas, United States.

[*] Corresponding author: Bo Li, E-mail: libo@tamu.edu.

**Abstract**

Despite significant anecdotal evidence regarding the vulnerability of the U.S. power infrastructure, there is a dearth of longitudinal and nation-level characterization of the spatial and temporal patterns in the frequency and extent of power outages. A data-driven national-level characterization of power outage vulnerability is particularly essential for understanding the urgency and formulating policies to promote the resilience of power infrastructure systems. Recognizing this, we retrieved 179,053,397 county-level power outage records with a 15-minute interval across 3,022 US counties during 2014-2023 to capture power outage characteristics. We focus on three dimensions—power outage intensity, frequency, and duration—and develop multiple metrics to quantify each dimension of power outage vulnerability. The results show that in the past ten years, the vulnerability of U.S. power system has consistently been increasing. The national cumulative user outage time reached 7.86 billion user-hours, with a mean of 2.55 million user-hours at the county level, highlighting a significant disruption to customer service. Counties experienced an average of 999.4 outages over the decade, affecting an average of more than 540,000 customers per county, with disruptions occurring approximately every week. Coastal areas, particularly in California, Florida and New Jersey, faced more frequent and prolonged outages, while inland regions showed higher outage rates. A concerning increase in outage frequency and intensity was noted, especially after 2017, with a sharp rise in prolonged outages since 2019. The research also found positive association between social vulnerability and outage metrics, with the association becoming stronger over the years under study. Areas with higher social vulnerability experienced more severe and frequent outages, exacerbating challenges in these regions. These findings reveal the much-needed empirical evidence for infrastructure owners and operators, policymakers, and community leaders to inform policy formulation and program development for enhancing the resilience of the U.S. power infrastructure.

**Introduction**

Power infrastructure systems are crucial for the functionality and well-being of societies (Xu et al., 2024); however, the increasing frequency of extreme weather events combined with aging infrastructure are making power systems more vulnerable to disruptions (Casey, Fukurai, Hernández, Balsari, & Kiang, 2020; Mukherjee, Nateghi, & Hastak, 2018). Although recent studies on power outages provide anecdotal evidence regarding the vulnerability of power infrastructure in the United States, there is a lack of empirical support for the extent to which power outages evolving at the national level. Some studies analyzed certain blackout events under extreme weather conditions, such as the 2021 Texas power crisis caused by winter Storm Uri (Flores et al., 2023; Lee, Maron, & Mostafavi, 2022). Other studies evaluated the extent and distribution of climate change-induced power outages in a certain area or timespan. For example, Flores et al. (2024) characterized the impacts of storms on power outage in New York state during 2017-2020. In a recent national study, Do et al. (2023) characterized the spatiotemporal distribution of power outages under weather events across U.S. counties through 2018-2020. The study only focused on outages during extreme weather events omitting non-event-related outages. Moreover, current studies lack longitudinal assessment of power outages, which could impede monitoring the trend of power outage vulnerability in the U.S. The absence of empirical evidence regarding the spatial and temporal characteristics of power outage vulnerability in the U.S. has contributed to an absence of urgency for policies and resilience investments needed to alleviate the vulnerability. Recognizing this gap, in this study, we use fine-grained data related to 179,053,397 county-level power outage records with 15-minute intervals across 3,022 U.S. counties during 2014-2023 to capture power outage characteristics.

In the examination of power outage characteristics, multiple indicators have been proposed in the prior literature. Some researchers measure power outage extent as the time period during which customers without power exceed a certain threshold (e.g.(Do et al., 2023; Flores et al., 2024)). Flores et al. (2023) applied power-out person-time, which is an interaction term between outage duration and affected customers. Nevertheless, metrics for measuring power outages are scattered, and there is currently no framework for systemic measurement of the extent of power outage vulnerability. To address this gap, inspired by hazard vulnerability assessment studies (e.g., Casey et al., 2024; Miao et al., 2024), we applied a framework to assess power outage vulnerability based on three dimensions: frequency, duration, and intensity (Fig.1). Under each dimension, we developed multiple metrics to quantify and perform spatial and temporal analysis regarding the characteristics of power outage trends in the U.S.

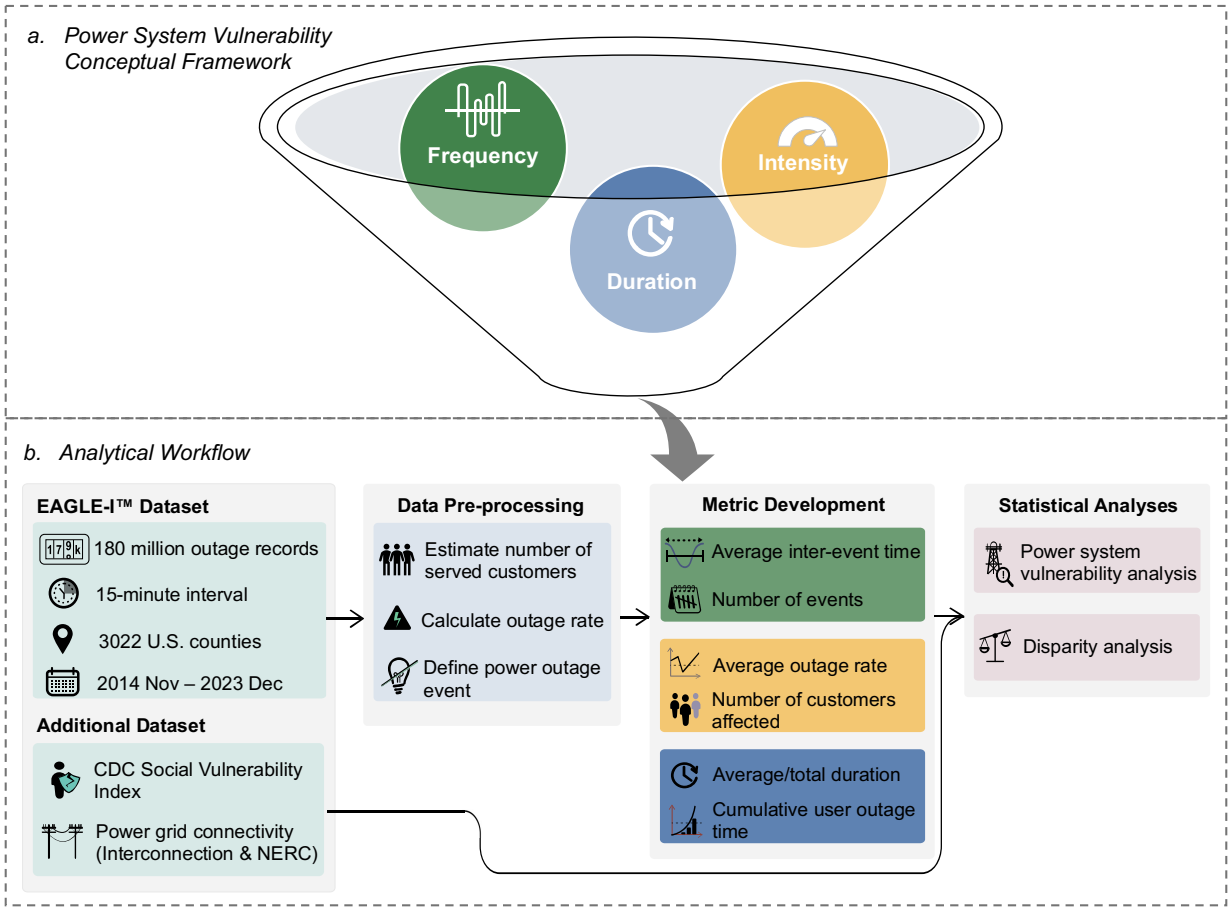

**Fig. 1 Power outage assessment metrics.** We assessed power outages from the three dimensions—frequency, duration, and intensity—and developed metrics for each dimension. See Methods for metrics definition.

The analysis used a large-scale, longitudinal, and high-spatiotemporal-resolution power outage dataset (Brelsford et al., 2024) to capture the characteristics of power outage trends in the U.S. We analyzed 179,053,397 historical power outage records across 3,022 counties with 15-minute temporal resolution from November 2014 through December 2023 from the Environment for Analysis of Geo-Located Energy Information (EAGLE-I™) platform (Energy). We assessed power outages using the metrics shown in Fig. 1 to capture the spatial and temporal trends in the outage metrics. Then, we examined whether socioeconomic and power grid connectivity disparities exist in power outage exposures and observed the trend over the past decade. The findings reveal multiple novel insights regarding the power outage vulnerability trends and characteristics in the U.S.: First, the analysis of power outage data from 3,022 U.S. counties between 2014 and 2023 revealed significant disruptions in the nation's power supply. On average, counties experienced an overall average of 999.4 power outage events over the decade, affecting over 540,000 customers per county. These outages occurred approximately every week, with counties experiencing power loss for about 3.65% of the time on average over the past decade. The national cumulative user outage time reached 7.86 billion user-hour, with a mean of 2.55 million user-hours at the county level, highlighting a significant service disruption to customers. Second, distinct spatial patterns emerged from the data, with coastal areas, particularly California, Florida, and New Jersey, experiencing more frequent and prolonged outages compared with inland regions. While coastal areas had higher numbers of affected customers due to greater population density, non-coastal states exhibited a higher outage rate, indicating a greater proportion of people affected by power disruptions. Third, the analysis identified a concerning trend of increasing power outages over the past decade, with a notable surge in events and affected customers after 2019. This trend highlights the growing vulnerability of the U.S. power system. The probability of experiencing extended outages (over 5% of the year) has risen sharply, especially after 2019. The geographical spread of "hotspots" — areas consistently experiencing prolonged outages — has expanded significantly.

Fourth, comparing the periods 2014-2018 and 2019-2023, the study found a significant increase in power outage intensity, with a higher average outage rate and a growing number of affected customers. Florida, New Jersey, Connecticut, and Washington D.C. experienced notable spikes in outage intensity, while California, Florida, and Texas consistently had the largest number of affected customers. Fifth, the analysis of outage frequency revealed an overall increase in events across all intensity and duration categories, with significant spikes after 2019. Interestingly, the relative proportion of outages in each category remained stable, suggesting proportional growth across different types of outages (in terms of extent and duration). The study also found that intervals between outages have shortened in recent years, meaning more frequent power disruptions. Finally, the research explored the relationship between power outage metrics and social vulnerability using the Social Vulnerability Index (SVI)(Agency for Toxic Substances and Disease Registry). Counties with higher social vulnerability were found to experience more frequent, longer, and larger-scale power outages, creating "dual burden" areas where social and power system challenges compound each other. This was particularly evident in California, Texas, Louisiana, and Florida. The study also highlighted disparities between urban and rural areas, with urban regions experiencing more frequent but shorter outages, while rural areas faced less frequent but longer and broader outages.

In sum, the findings offer several significant contributions to the understanding of power outage vulnerability in the United States. First, it provides robust empirical evidence regarding the extent, spatial variation, and temporal escalation of these vulnerabilities. Diverging from the predominantly event-specific analyses in existing literature, this research provides a longitudinal and national-level perspective, shedding light on the magnitude, hotspots, and evolution of community susceptibility to power outages across the country. These insights inform interdisciplinary researchers in engineering, disaster studies, and geographical sciences, informing studies on community vulnerability, energy justice, and infrastructure resilience. Furthermore, the spatiotemporal characterization of power outage vulnerability underscores the urgency of addressing this issue at national, state, and local levels. These insights are crucial for power infrastructure owners and operators, policymakers, and community leaders as they develop policies and implement strategies aimed at prioritizing infrastructure and optimizing resource allocation to enhance power systems resilience.

**Results**

**Overall spatiotemporal trends of power outages in the U.S. from 2014 to 2023**

This study first summarized the statistics of six power outage metrics spanning from November 2014 to December 2023 in 3,022 counties of the contiguous United States (CONUS) (Table 1). Across the study period, U.S. counties experienced 999.4 power outage events on average (IQR: 486.25, 1369; max: 5761). The average total power outage duration in ten years is 118.79 days (IQR: 53.61, 155.90; max: 890.25), which means each US county has experienced a power outage 3.65% (IQR: 1.71%, 4.69%; max: 25.96%) of the time over the past decade. The average interval between power outage events is 7.16 days (IQR: 2.22, 5.71; median: 3.37), indicating that the counties experienced a power outage event approximately every week. The mean power outage rate is 1.51% (IQR: 0.70%, 1.54%; median: 1.02%). Over the past ten years, power outage events cumulatively affected a mean of 540,915 customers in each county (IQR: 61,962, 486,026; median: 191, 557), indicating a substantial and widespread impact. The cumulative user outage time reaches 7,863,993,105 user-hours, with a mean of 2,548,280 user-hours at the county level (IQR: 169,535, 2,030,562; median: 636,909), highlighting a significant disruption to customer service, which could cause significant impact on daily life and economic activities in each county.

Table 1 Summary statistics of the six power outage metrics.

| Metrics | Mean | Max | Min | Median | Interquartile range (Q1, Q3) | Standard Deviation |
|---|---|---|---|---|---|---|
| Number of events | 999.41 | 5761.00 | 1.00 | 888.00 | (486.25, 1369.00) | 711.83 |
| Average outage rate (%) | 1.51 | 100.00 | 0.10 | 1.02 | (0.70, 1.54) | 0.03 |
| Total duration (days) | 118.79 | 890.25 | 0.01 | 94.69 | (53.61, 155.90) | 99.23 |
| Average Inter-event time (days) | 7.16 | 273.51 | 0.00 | 3.37 | (2.22, 5.71) | 16.46 |
| Number of customers affected | 540914.72 | 32050993 | 2 | 191556.50 | (61961.75, 486026.25) | 1318285.38 |
| Cumulative user outage time (hours) | 2548280.33 | 137422361.5 | 0 | 636908.75 | (169534.75, 2030561.69) | 7027485.95 |

We observed distinct spatial patterns for the power outage metrics (Fig. 2). U.S. coastal areas, including the West Coast, East Coast, and Gulf of Mexico suffered more severely from power outages, with greater power outage frequency and longer duration. California, Washington, Maine, New Hampshire, Massachusetts, New Jersey, and Florida, experienced power outages with an interval of less than two days, while the interval for most counties in Minnesota, Iowa, and North Dakota is greater than eight days (Fig. 2d). Along with the factor of denser population, coastal areas have a much higher number of affected customers and longer user outage time. However, the average power outage rate for non-coastal states is higher than the coastal areas, indicating a higher proportion of people living in non-coastal states suffered from power system disruptions compared with those residing in coastal states.

We also identified temporal trends of power outages over the ten years (Fig 2). For example, the yearly number of outage events has kept increasing since 2015 (Note that we only retrieved outage data of 2014 for two months, so we didn't take the year 2014 into consideration when comparing most metrics to avoid potential bias.). The number of outage events surged after 2017, with a rate of approximately 30%. From 2018 to 2023, the yearly number of outage events continued to grow at a relatively slower but steady rate. Total duration, average inter-event time, and the number of affected customers show a similar increasing trend, with a surge after 2016 (2019 for average inter-event time) and remained high during the following years. The average outage rate and cumulative outage time does not show a clear temporal pattern, which needs further analytics to identify. Overall, U.S. residents experienced more frequent power outages and endured longer outage duration over the past decade, indicating the increasing vulnerability of the power system.

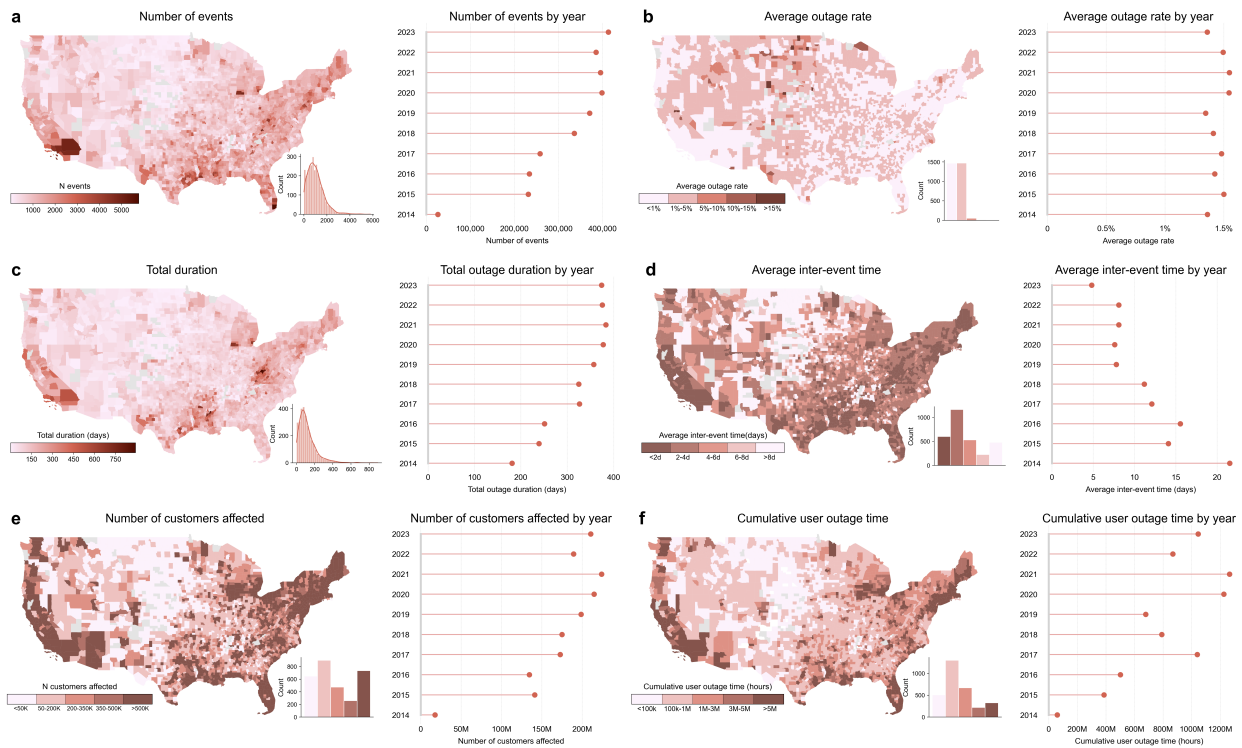

**Fig. 2 Six metrics of county-level power outage data from 2014 to 2023.** a. Number of outage events; b. Average outage rate; c. Total duration; d. Average inter-event time; e. Number of customers affected; f. Cumulative user outage time. Maps display the spatial distribution of ten-year summarizing statistics of the metrics. Plots on the bottom right corner of the maps show statistical distributions. The lollipop plots on the right side of each sub-figure show the metric year by year. Maps include 3,022 counties, with gray areas indicating counties without data.

**Characterizing U.S. power outages regarding duration, frequency and intensity**

To get a more nuanced understanding of power outage characteristics, we examined the spatial and temporal trends for each metric in detail. For power outage duration, we calculated the yearly total outage duration of each county and then normalized the indicator by dividing the length of the year. The normalized indicator is called outage duration proportion, measuring the percentage of time in a year that people experience power outages. Fig. 3a displays distributions of yearly outage duration proportion. Comparing on a year-to-year basis, the peaks of the probability density plots keep moving right, suggesting a stable increasing trend of power outage duration. In the year 2014, the probability that the duration proportion exceeds 5% is 0.069 (95% CI: [0.060, 0.078]), while the probability grew to 33.3% (95% CI: [0.315, 0.349]) in 2023. The sharp increase indicates that compared to ten years ago, U.S. residents now are more likely to experience 5% of a year's time without power. The occurrence of outage proportion exceeding 10%, namely more than 36.5 days without power cumulatively in a year, was scarce in 2014, with the probability of only 2% (95% CI: [0.015, 0.026]), while the probability has risen to 8.3% (95% CI: [0.074, 0.094]. The higher probability of experiencing extended power outages suggests that the reliability of the U.S. power system has deteriorated. Table SI 1 in the supplementary information shows the outage proportion threshold and corresponding probabilities for every year. From the table, we can observe the year-to-year variability of total outage duration. We also noted a surge in the extended outage probability around the year 2019, so we divided the data into two periods: before and after 2019. Fig. 3b compares the distribution plot of outage duration proportion for the two groups. Total outage duration in 2019-2023 is significantly higher than that in 2014-2018 ($p < 0.05$). The cumulative probability distribution function suggests a similar trend. The curve for 2014-2018 is consistently steeper than the curve of the latter period, illustrating that for most counties, outage duration is relatively short in this period. The flatter curve for 2019-2023 shows that counties are more likely to experience prolonged outages in the past five years. Comparing the two groups, we can find that power outages have consumed a larger fraction of the time during the most recent five years, reflecting a clear decline in power system reliability from 2014-2018 to 2019-2023. We calculated the 5-year average of outage proportion for each county and labeled them into four categories: severe (duration proportion larger than 10%), major (duration proportion between 5% and 10%), moderate (duration proportion between 1% and 5%) and minor (duration proportion smaller than 1%). Fig. 3c dissects the relative share of counties in each category and makes comparisons between years 2014-2018 and years 2019-2023. The results show that the number of counties falling in severe, major, and moderate categories are rising in the most recent five years. For example, 26% of the counties (414 of 1756 counties) in the moderate category elevated into the major category, and 27% of counties (101 of 370 counties) with major labels moved to the severe category. For counties from minor to moderate category, the percentage reaches 60% (373 of 620 counties). In general, power outage duration for counties tend to increase to various extents in the most recent five years. As a comparison, there are fewer counties with shortening power outage duration. Furthermore, we made investigations on the counties that repetitively experience long power outages over the years. We set four thresholds on outage duration proportion (Fig. 3d and Fig. 3e), and then defined hotspots as counties where the yearly accumulative outage duration exceeds these thresholds every year during the research period. For example, Kanawha County in West Virginia is the only county with ">15%" label during 2014-2018, which means Kanawha County was without power cumulatively for more than 54 days each year. In this way, we identified areas that consistently experiencing prolonged outages over the years. During 2014-2018, those hotspots are mainly found in several geographical regions, such as the northeastern areas, as well as parts of Florida, Louisiana, Texas, and California. During 2019-2023, however, the geographical range of hotspots expanded significantly, covering nearly the entire continental U.S. except for the central states. The spread of hotspots across almost the U.S. suggests that the increasing vulnerability of power systems is becoming more pervasive and widespread. The repetitively affected areas are no longer confined to a few regions but are spreading to new areas. Numerically, the number of hotspots, across all thresholds, has increased multiple times, which means more areas consistently experiencing extended disruptions in the electricity supply.

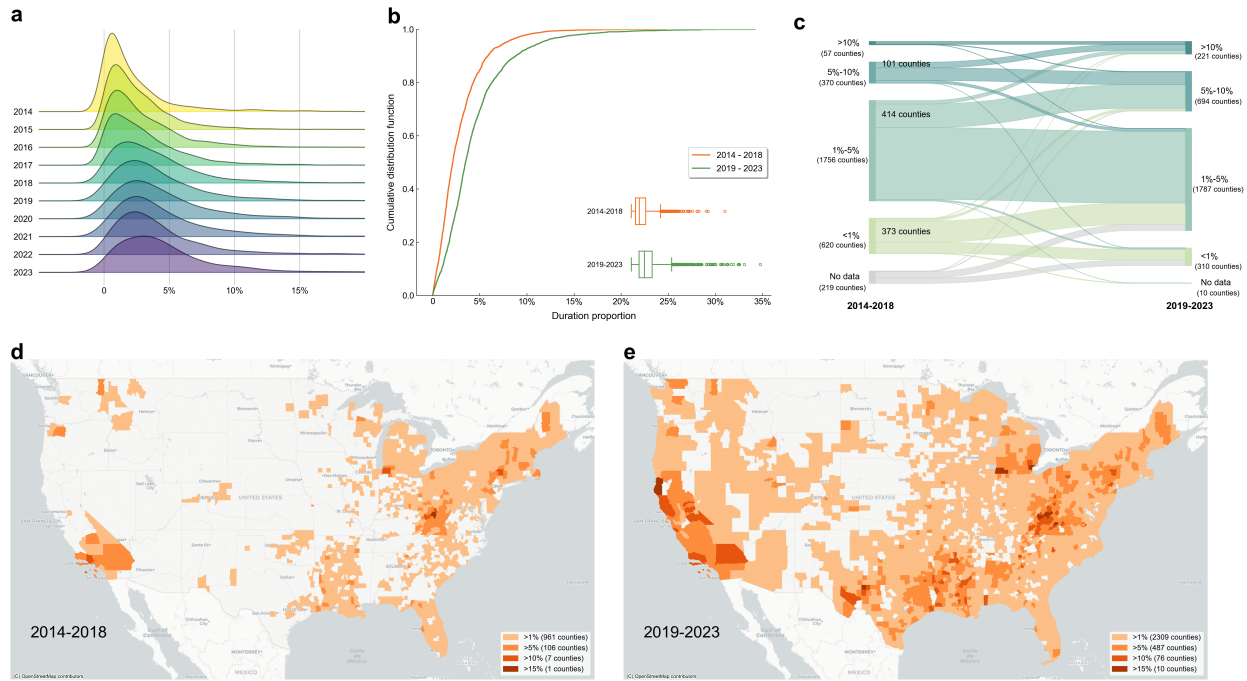

**Fig. 3 Descriptive statistics of power outage duration.** a. Probability density plots of year-by-year outage duration proportion; b. Distribution plots of outage duration for US counties during the period 2014- 2018 (n = 2803) and the period 2019-2023 (n = 3012). The curves show the cumulative distribution function. The Mann-Whitney U test was performed to examine group differences ($p < 0.05$); c. Sankey plot displaying how the outage proportion changes over the ten years for each county; d. Spatial distribution of hotspots with repetitively prolonged power outages during 2014-2018. e. Spatial distribution of hotspots with repetitively prolonged power outages during 2019-2023.

We examined power outage intensity based on the power outage rate and the number of affected customers. To track the trend of power outage magnitude over the years, we calculated county-level average outage rate across all outage events that occurred during periods 2014-2018 and 2019-2023 separately. To facilitate comparison, we sorted the counties in descending order of average outage rate for each period and compared the outage rate values for counties in the same position. Results (Fig.4a) show that the average outage rate during 2019-2023 is significantly higher than that during 2014-2018 (paired Wilcoxon test, $p<0.05$), demonstrating that power outages have intensified in the past five years. Another indicator we considered is the number of customers affected by outage events, as this measure directly reflects the broad impact that power outages can have on communities. Fig. 4b captures the temporal trend of the peak number of affected customers over ten years. The peak refers to the maximum number of customers impacted across all outage events each year. We calculated this metric at the county-level and then computed the yearly average value for the entire continental United States. Despite some fluctuations, the overall trend for average peak number of customers is ascending (upper plot of Fig. 4b). Considering that the total number of customers keeps almost constant over the years (less than 5% increase), we computed the relative proportion for peak number of affected customers and identified an even more notable increase as well (lower plot of Fig. 4b). The upward trend in both the absolute number and the relative share across the study period highlights a substantial growth in the scale and intensity of these outage events. Fig. 4c depicts the spatiotemporal trend of power outage intensity at the state level. Florida, New Jersey, Connecticut, and Washington D.C. have significant spikes, indicating that those regions have been impacted by large-scale blackouts in certain years. The lower plot of Fig. 4c displays the hotspots of the yearly accumulated number of affected customers, which is the sum of affected customers in a year. Spatially, California, Florida, and Texas have the largest number of affected customers. Temporally, we can also identify the year-to-year viability of outage intensity and the years with the most severe impacts. Those plots enable a better comparison of outage impacts across different regions and time periods.

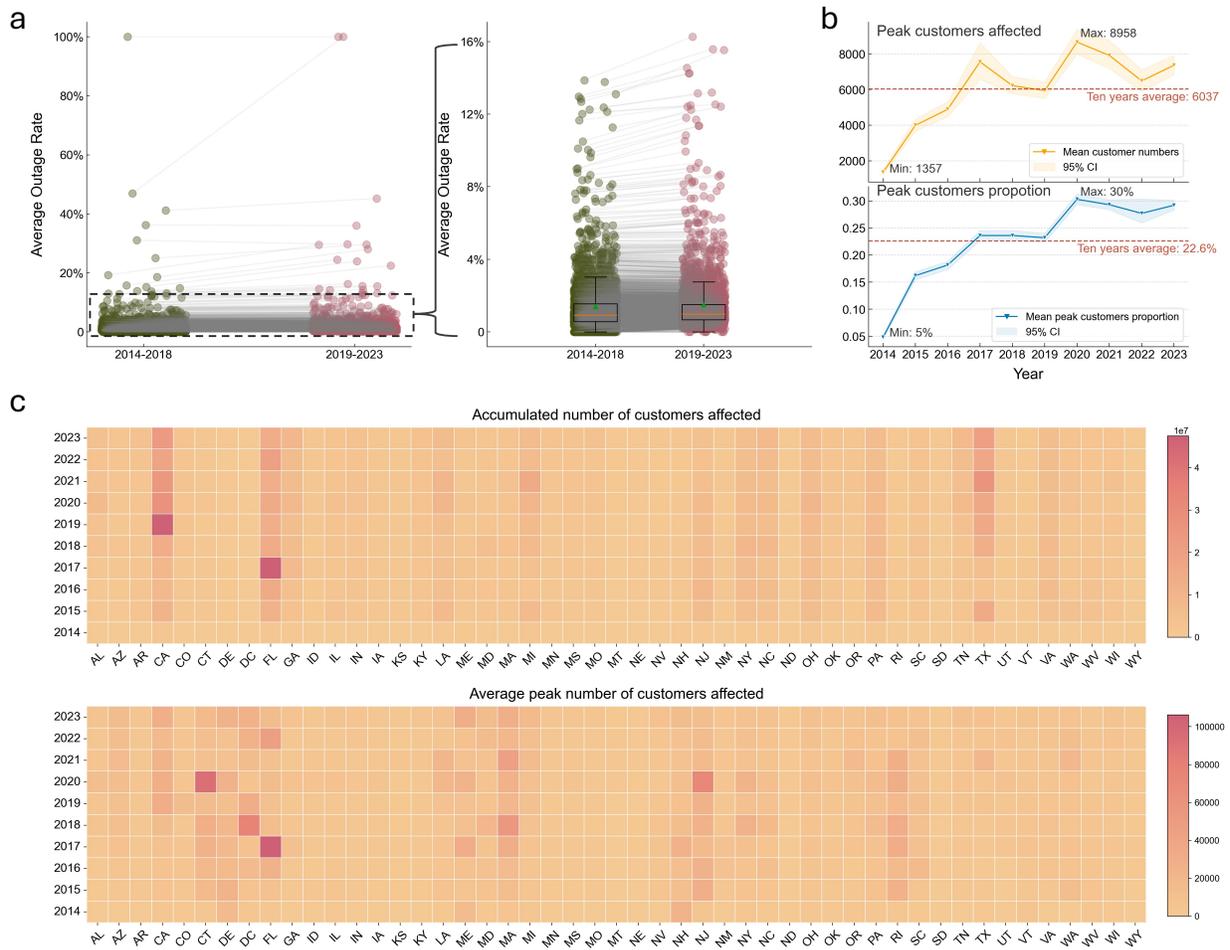

**Fig. 4 Descriptive statistics on power outage intensity. a.** The line-linked paired points represent the average power outage rate of counties in the same position during 2014-2018 and 2019-2023 periods, respectively. Plots with average outage rate in the range (0%, 16%) are zoomed in. The 2019-2023 outage rate is significantly higher than that of 2018-2023 (Paired Wilcoxon test, $p < 0.05$), indicating an upward trend in the average outage rate. Boxplot shows the distribution of outage rate values, with a median line, mean triangle, and box ends representing first and third quantiles. Whiskers extend to values within 1.5 times the interquartile range. **b.** Line plots for the peak number of affected customers by year. The upper plot shows the absolute number of affected customers, while the lower plot shows the relative share regarding the total number of customers. Maximum and minimum values are denoted. The shaded area shows a 95% confidence interval of the line plot. **c.** Heatmap for the number of affected customers. The X-axis shows the US states.

In analyzing power outage frequency, we categorized power outage events with varying intensity and duration categories and counted the number of outage events in each category. Fig. 5a displays the trend of event numbers for all intensity categories over the ten-year study period. Overall, the event numbers are increasing with notable spikes after year 2018. A similar trend can be observed from Fig. 5b, which denotes the trend related to the event numbers for events with various duration. We examined the relative proportion of all event categories (Fig.5c and 5d). Over 80% of the outage events have a power outage rate lower than 10%, indicating that less severe power outages are more common. However, events with an outage rate over 50% account for approximately 4%, suggesting that the risk of extensive outage is still a notable concern. Events with a duration of one to two hours are the most frequent, accounting for 60%, and events lasting longer than two hours are much less. Interestingly, we found that the relative proportion of all event categories remained steady over the ten years. For example, events with a duration between 0.5 to 1hour account for 23.2% of outages with a small standard deviation of less than 1%. Table SI 2 and SI 3 in the supplementary information shows the average proportion and standard deviation for all event categories. This finding reveals that the growth for outage events with varying intensity and duration is proportional.

Another metric we applied to measure power outage frequency is inter-event time. We examined the distribution of inter-event time for periods 2014-2018 and 2019-2023, separately. The Kolmogorov-Smirnov (K-S) test statistic is 0.02 for the period 2014-2018 and 0.01 for the period 2019-2023, meaning that both groups exhibit good power law fit. We also employed loglikelihood ratio to exclude other distributions. Prior studies suggest that outage interval follows exponential distribution (Carreras, Newman, & Dobson, 2016; Chen, Thorp, & Parashar, 2001). However, compared to exponential distribution, the log-likelihood ratio for period 2014 -2018 is 2814.2 with $p < 0.05$, and 1344.2 with $p < 0.05$ for the period 2019-2023, suggesting power law is a better fit. As power law distribution usually emerges from complex systems with interconnected components and cascading failures, the fitted distribution suggests that power systems are complex networks in which local failure can trigger cascading effects, leading to large-scale disruptions. Power law fit for outage inter-event time indicates that the occurrence of outages is not random, but rather follows a specific statistical pattern. In this way, the fitted power law distribution can assist with the outage occurrence prediction. Fig. 5e displays the fitted parameters of power law distributions. Distributions for the two groups are statistically different (K-S test, $p < 0.05$). The fitted parameter alpha for both groups suggests a heavier tail for the period 2014-2018, indicating that longer intervals between outage events are more likely during that time. In other words, the intervals between outage events tend to become shorter, resulting in more frequent occurrence of power disruptions in the most recent five years.

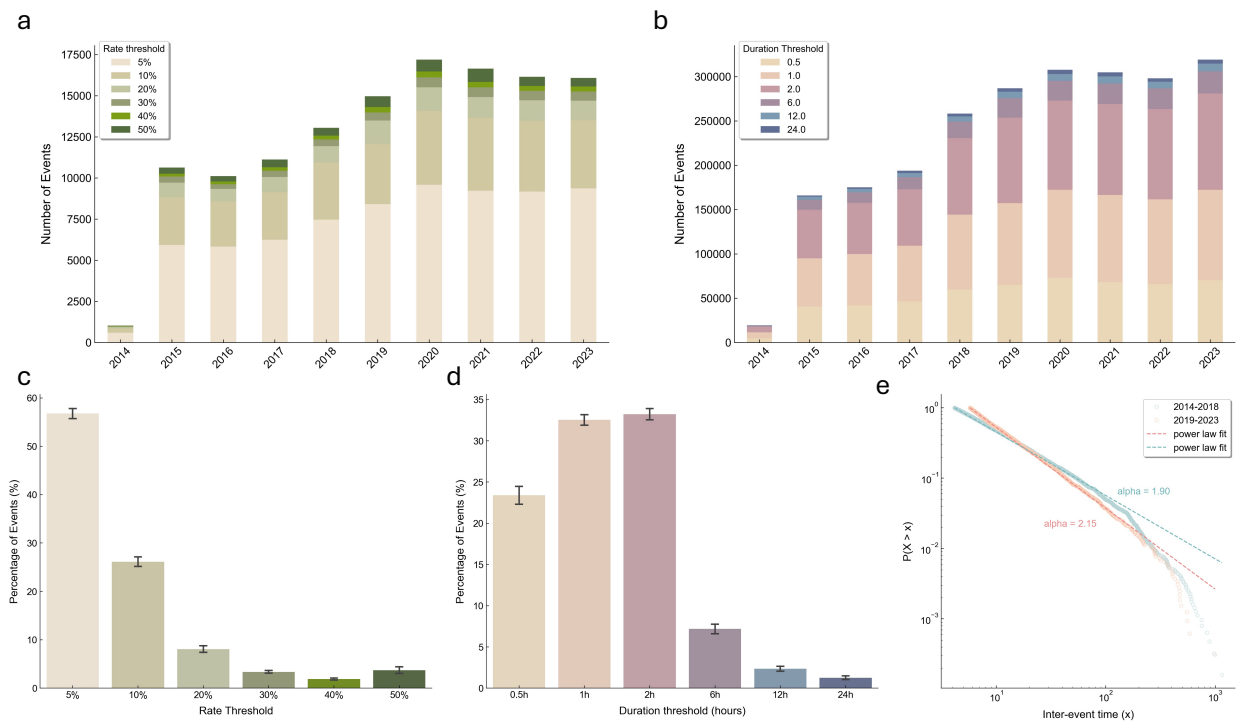

**Fig. 5 Descriptive analysis of power outage frequency. a.** Stacked bar plot for event numbers with varying outage intensity. **b.** Stacked bar plot for event numbers with varying outage duration. **c.** Relative proportion for all event intensity categories. **d.** Relative proportion for all event duration categories. The error bar denotes a 95% confidence interval. **e.** Cumulative probability distribution of inter-event time for periods 2014-2018 and 2019-2023.

**Capturing unequal power outage distribution regarding socioeconomic status and geographical regions**

We used the social vulnerability index published by US Centers for Disease Control and Prevention and Agency for Toxic Substances and Disease Registry (CDC/ATSDR) as a comprehensive indicator for socioeconomic status and then examined the relationship between power outage metrics and SVI. We categorized counties into three groups based on the tertiles of SVI, and labeled groups as low, medium, and high social vulnerability. Similarly, we categorized and labeled counties with tertiles of every power outage metric: the number of events, outage rate, outage duration, and inter-event time. Fig. 6 shows the geographical distribution of areas with varying extent of social and power system vulnerability for outage duration. Figures for the other outage metrics are available in supplementary information (Fig. SI 1-3). We defined counties with high social vulnerability and severe power outages as counties with dual burdens since those counties are vulnerable to both social and power system challenges that can exacerbate their difficulties during crises. From the figures, we can observe that the geographical range of counties with dual burdens is largely expanding, especially in California, Texas, Louisiana, and Florida.

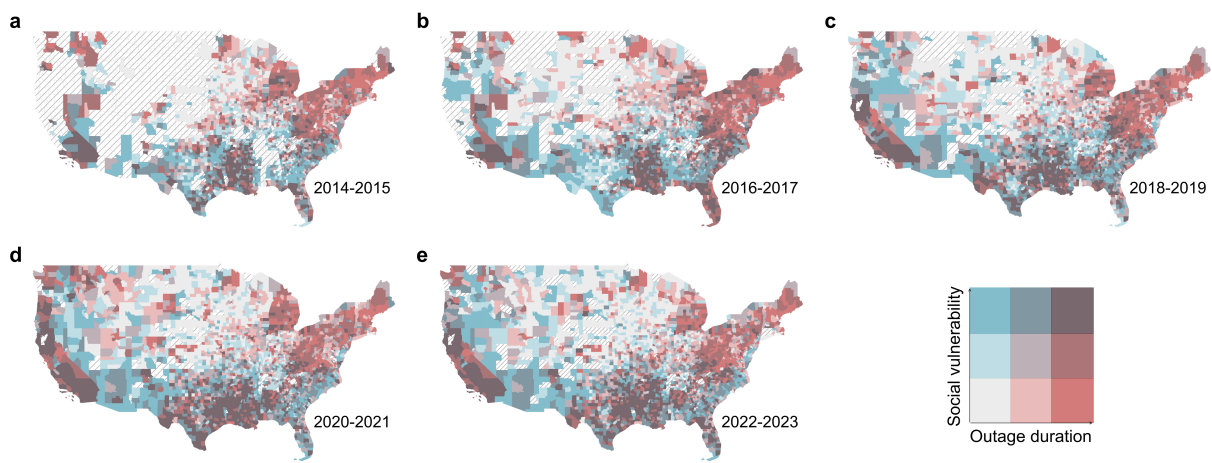

**Fig. 6 Geographical distribution of counties with dual burdens from 2014 through 2023.** SVI is updated every two years, so we calculated a two-year average for outage metrics to create the bivariate maps. The X-axis of the legend displays short, medium and long outage duration groups from left to right; the y-axis displays low, medium, and high social vulnerability groups from bottom to top. The upper-right grid of the legend highlights counties with dual burdens.

Further, we performed an ordinary least square (OLS) regression to further assess the relationship between SVI and power outage metrics. Results show significant positive associations between SVI and all the outage metrics (Fig. 7), illustrating that population groups with higher social vulnerability tend to suffer from outages with higher frequency, longer duration, and larger scale. The association suggests that systematic inequality may exacerbate the adverse effects of power outages on socially disadvantaged groups, causing significant hardship and negative well-being effects and even potentially leading to a cycle of vulnerability. Moreover, OLS regression was performed every two years from 2014 through 2023, and the coefficients keep an increasing trend over the years. This trend suggests that the disparity in outage metrics across populations with varying degrees of vulnerability has been widening over the past decade. The growing disparity is noteworthy, as this reflects that infrastructure maintenance and upgrade policies may not adequately address equity in infrastructure prioritization and resource allocation over the years.

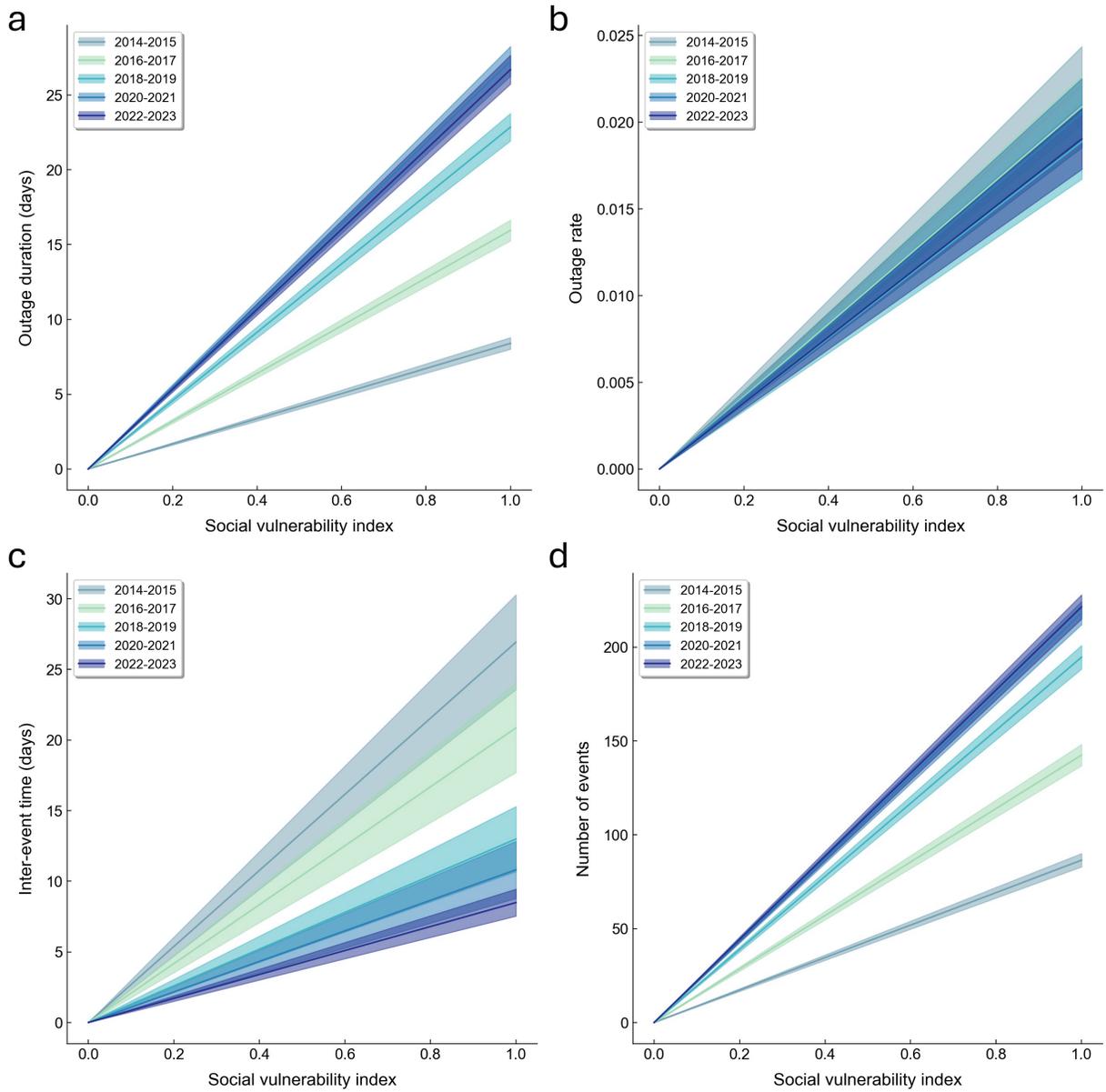

**Fig. 7 OLS regression between SVI and outage metrics. a.** OLS regression between SVI and outage duration. **b.** OLS regression between SVI and outage rate. **c.** OLS regression between SVI and inter-event time. **d.** OLS regression between SVI and number of events. The shaded area denotes a 95% confidence interval of the plot.

The analysis results related to the counties with dual burdens demonstrate a similar temporal trend. Fig. 8 shows that counties with dual burdens have experienced more severe outages over the years, characterized by a greater number of power disruptions, prolonged time without power, larger scale, and shorter outage intervals. This observation indicates a compounding effect where both social and power system vulnerability interact with each other, leading to disproportionately severe impacts on the residents in these counties. Power system failure may amplify the negative effects of issues, such as housing instability, unemployment, limited access to resources (Dargin & Mostafavi, 2022), and raising more health and safety concerns. The worsening trend highlights the urgent need for targeted policies and programs to enhance the robustness and resilience of the power system.

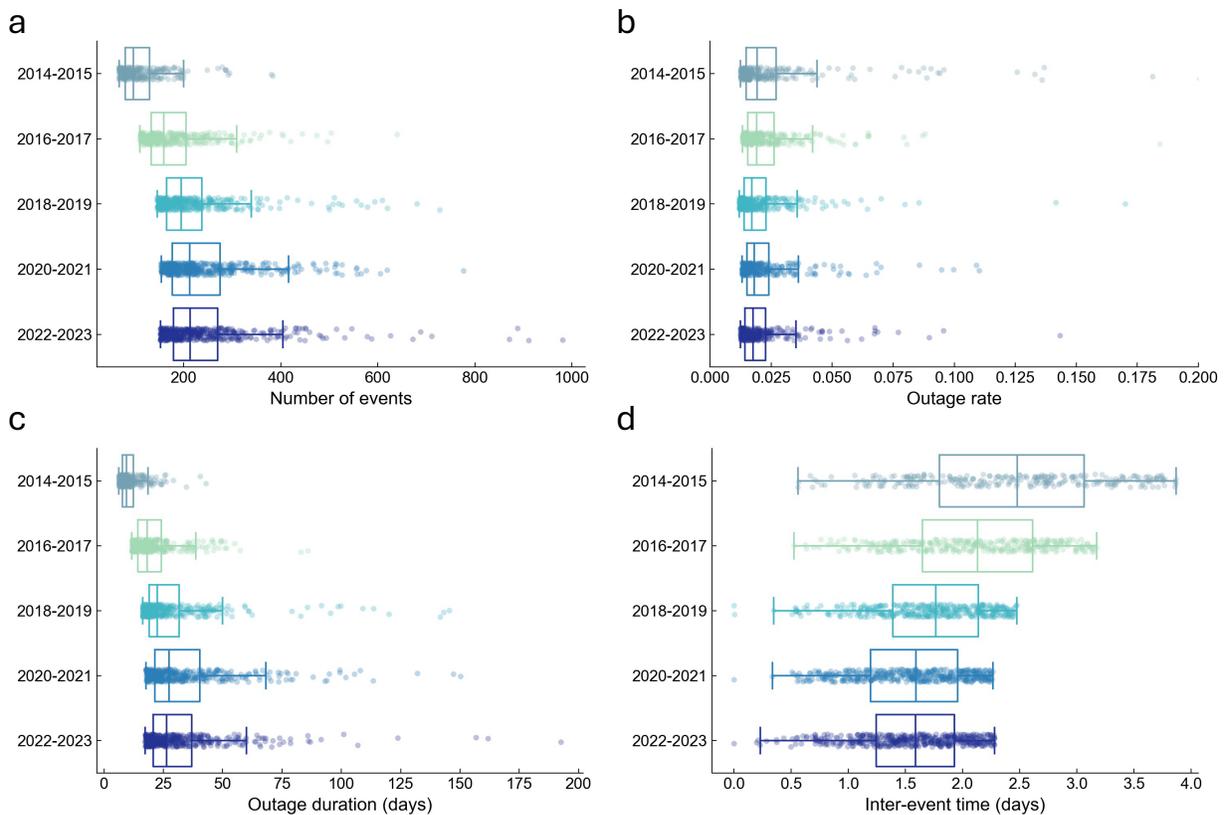

**Fig. 8 Distribution plots of outage metrics over the past ten years for counties with dual burdens. a.** Boxplot for number of events. **b.** Boxplot for outage rate. **c.** Boxplot for outage duration. **d.** Boxplot for inter-event time. The Kruskal-Wallis H test was employed to determine the significance of differences among the five groups ($p < 0.05$).

We also examined the relationship between power outage metrics and urbanicity. Counties were grouped into either urban or rural categories and compared between the two categories. Notably, urban and rural areas take on opposite patterns regarding the characteristics of power outages.

Urban areas have a higher number of outage events and shorter intervals between events for every year during the study period, while the average outage rate is lower than urban regions (Fig.9a-9c, Man-Whitney U test, $p < 0.05$ for all groups). The differences in the outage duration are significant for years 2015, 2021, and 2022 during which the average outage event duration of urban areas is slightly shorter (Fig. 9d, Man-Whitney U test, $p < 0.05$). From the analysis, we can summarize that urban outage events are characterized as high frequency, short-term disruption, and small scale. In contrast, rural power outage events are less frequent, but once they occur, they may have broader impacts and last longer durations. The distinct patterns of outage characteristics provide insights into the unique challenges faced by rural and urban areas, respectively. In urban areas, the higher frequency of outages can be attributed to the dense development and co-location of power grid to the built environment and the resulting increased likelihood of cascading incidents, such as equipment failures, accidents, and construction activities. Rural areas, however, should focus on enhancing the proximity of power restoration efforts to avoid long-lasting downtime and the following adverse effects.

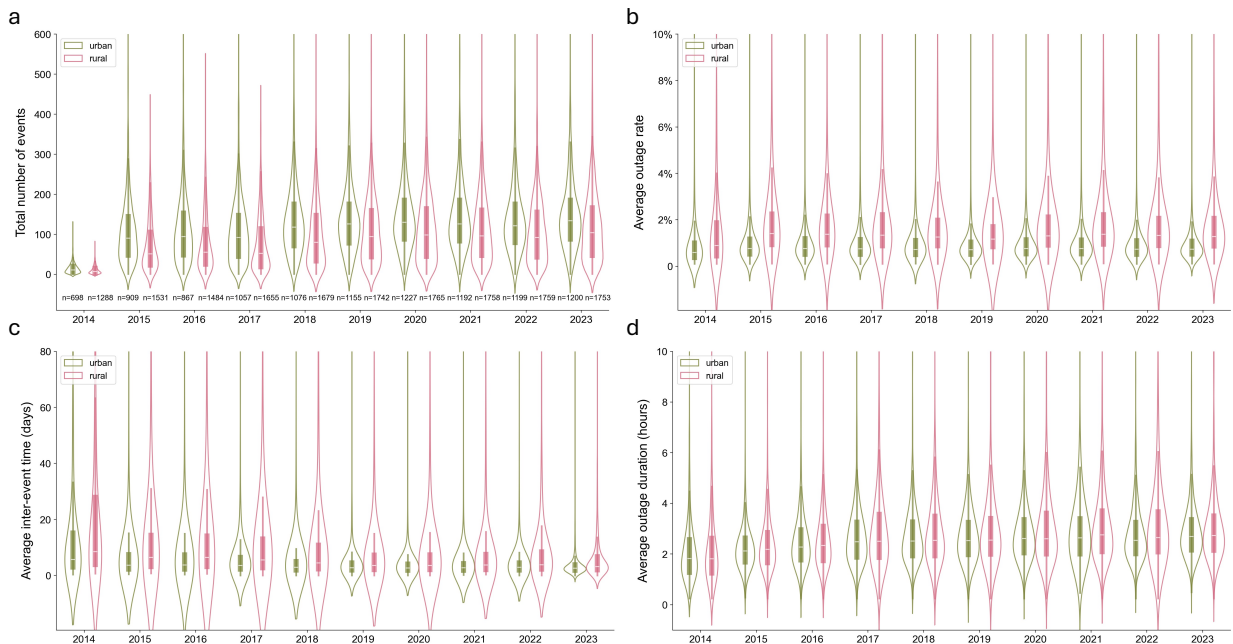

**Fig.9 Urban-rural differences regarding power outage metrics. a.** Total number of events is higher in urban regions compared to rural regions. Numbers at the bottom of the plot denote county numbers for each group. The numbers are the same for Fig. 9b, 9c and 9d. **b.** The average outage rate is lower in urban regions. **c.** The average inter-event time is shorter in urban regions. **d.** The average outage duration is longer in rural regions. The Mann-Whitney U test was performed to examine group differences ($p < 0.05$ for all groups except for average outage duration of year 2014, 2016-2020 and year 2023).

To further evaluate geographical disparities in power outage metrics, we divided counties according to power grid connectivity data. The North American grid is composed of three major parts: Eastern Interconnection, Western Interconnection, and Electric Reliability Council of Texas. As cases are possible that counties belong to multiple interconnections, we labeled those counties as "boundary". Fig. 10a shows the relative share and temporal changes in outage duration of each interconnection category. Assuming all counties experience an equal length of time without power, we can obtain a baseline showing the situation in which no disparities exist. In Fig. 10a, the relative share of total outage duration for the Eastern Interconnection is decreasing, while it is noticeably increasing for Texas and the Western Interconnection. This indicates that the outage duration experienced by customers in Texas and the West is growing significantly compared to the East. Since the Eastern Interconnections covers the largest proportion of counties (approximately 80%) we studied, we further divided those counties into different categories according to the reliability entities of North American Electric Reliability Corporation (NERC) they belong to. For most NERC regions, the relative outage duration remains stable, but the SERC Reliability Corporation region has seen an increase in the outage duration each year. SERC includes 870 counties in our study, encompassing Alabama, Georgia, Mississippi, Missouri, North Carolina, South Carolina, and Tennessee. Our examination of the spatial disparity between the two types of grid connectivity suggests that all U.S. residents are experiencing greater power systems vulnerability, but there is also significant spatial variability in vulnerability across regions.

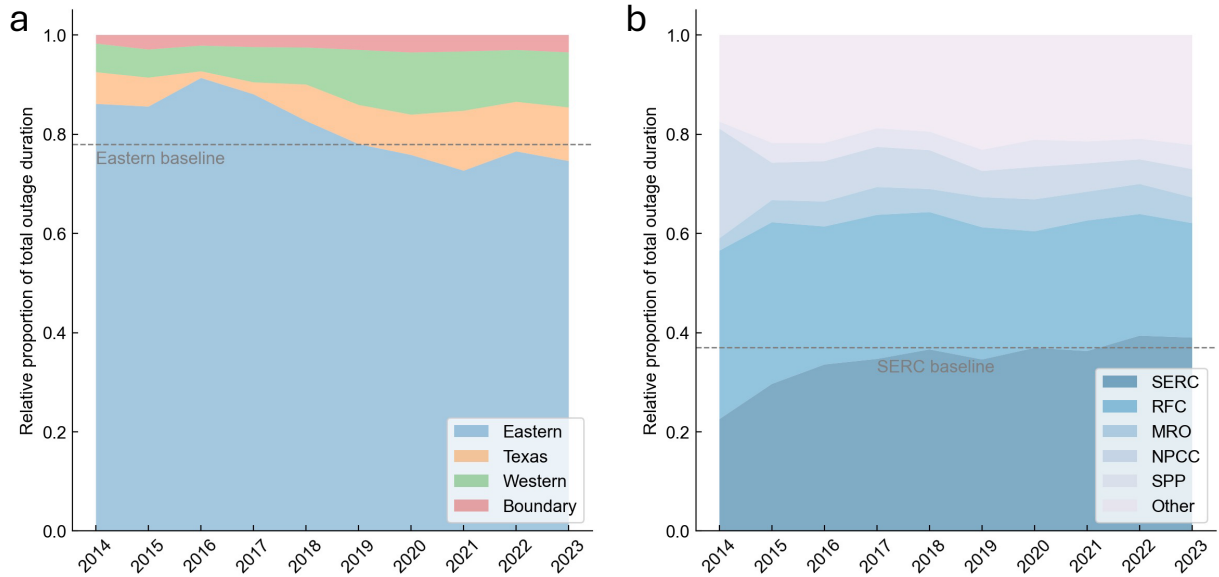

**Fig. 10 Changes of relative outage duration proportion among different grid connectivity areas. a.** Year-on-year changes in the share of relative outage duration proportion between different interconnections. **b.** Year-on-year changes in the share of relative outage proportion between different NERC regions. The baseline shows the situation in which no disparities exist by assuming all counties experience an equal length of time without power.

**Discussion**

Although there is considerable anecdotal evidence highlighting the vulnerabilities of the U.S. power infrastructure, comprehensive studies that systematically analyze these issues over time and across the nation are scarce. There is a critical need for a data-driven, nationwide analysis that characterizes the spatial and temporal trends in power outage frequency and impact. Such an analysis is essential for accurately assessing the severity of the problem, understanding the spatiotemporal characteristics, and informing the development of effective policies and programs aimed at enhancing the resilience of the U.S. power systems. This approach would provide the necessary foundation for strategic planning and investment to protect against future disruptions. Recognizing this, in this study, we retrieved 179,053,397 county-level power outage records with a 15-minute interval across 3,022 U.S. counties during 2014-2023 to capture power outage characteristics. We applied a standard vulnerability assessment model to examine power outages from intensity, frequency and duration dimensions. We developed multiple metrics for each dimension as measurements and then utilized the metrics to characterize the spatial and temporal trends. On average, counties experienced nearly 1,000 power outage events over the ten-year study period. Each U.S. county spent a total of 118.79 days without power, accounting for 3.65% of the time in ten years. Power outage events occurred approximately every 7 days, indicating that counties were without power once a week. The average outage rate is around 1%; if multiplied by the larger number of power system customers, this seemingly small percentage translates into a substantial number of affected customers. The accumulative outage user-time reached 7.8 billion user-hours, indicating the vast scale of impacts on social activities and well-being. These results depict a sketchy, but alarming picture of how ubiquitous and widespread power outages have been across the country.

We performed further analyses on the power outage metrics across the three dimensions to reveal important spatial and temporal patterns. Outage metrics consistently show an increasing trend of U.S. power systems vulnerability over the past decade. Power outages in the U.S. are becoming prolonged, intensified, and more frequent. Comparatively, residents on average spent 2% of the time without power during 2014-2018, while the percentage rose to 8.3% during 2019-2023. The impact scale of outages can be inferred from both the growing average outage rate and the number of customers affected. For example, outages at their peak moment in 2015 on average left 15% of the customers without power, while peak moments of outages in 2020 increased to 30%. The number of outage events maintained an upward trend for every intensity and duration category. Following power law distribution, the event interval for 2019-2023 has a lighter tail compared with that of 2014 - 2018, meaning that shorter intervals between outages have become more likely over the past five years. The worsening trend for all power outage metrics underscores that U.S. power systems are faced with increasing vulnerability. Based on the data-driven national-level evaluations of the duration, frequency, and intensity of power outages, infrastructure owners, operators, and policymakers can better understand the urgency of taking action to improve the resilience of power infrastructure in the U.S.

The characteristics of power outages display certain spatial patterns. For example, areas repetitively affected by prolonged power outages are primarily located in coastal regions. The coastal states of California, Texas, and Florida are on the top list regarding the number of customers being affected by outage events. The spatial patterns also indicate the existence of potential disparities. We also observed distinct power outage characteristics between urban and rural regions. Power outages in urban regions are more frequent with lower intensity and shorter duration, while rural outages are less common but yield to larger impact scale. Spatial variabilities also exist among different power grid connectivity regions. Compared to the East Interconnections, the West Interconnections and Texas have accounted for a relatively larger share of outage duration in recent years, reflecting the greater challenges these regions face in maintaining grid stability.

Previous literature has noted that power outages disproportionately affect socioeconomically disadvantaged population groups such as low-income and minority groups(Coleman et al., 2023; Lee et al., 2022; Liévanos & Horne, 2017), most in weather-related power outage events. This study incorporated the social vulnerability index into ten-year national power outage data and identified positive association between SVI and outage metrics. This finding indicates that the unequal energy resilience is not confined to certain areas or events. Instead, it might be a systematic issue that socially vulnerable groups are continuously faced with more frequent, larger scale and prolonged power outages. More notably, the positive association is becoming stronger over the years, indicating an enlarging gap of outage disparities. Areas with high social vulnerability have been experiencing worsening power outage issues, such as longer time without power and shorter intervals between outage events. This increasing disparity highlights a concerning trend that power system resilience of socially vulnerable communities is further compromised.

The study findings offer multiple important contributions. The study provides empirical evidence revealing the extent, spatial variation, and temporal escalation of power outage vulnerability. Unlike event-specific studies, which are prominent in the literature, this study provides a longitudinal and national-level perspective on the magnitude, hotspots, and evolution of vulnerability to power outage for communities across the U.S. These outcomes provide empirical knowledge to inform interdisciplinary researchers across the fields of engineering, disasters, and geographical science about the extent of susceptibility of U.S. communities to power outages for community vulnerability, energy justice, and infrastructure resilience studies. Also, the spatiotemporal characterization of power outage vulnerability is crucial for understanding the urgency of the problem at the national, state, and local levels to inform power infrastructure owners and operators, policymakers, and community leaders in developing policies and implementing strategies for infrastructure resilience prioritization and resource allocation.

There were also limitations with the retrieved power outage vulnerability metrics. The county-level assessment may obscure sub-county-level trends due to increased development and population influx over the years. While we found a correlation between county-level outages and high SVI, this relationship might differ with finer-scale data. Future studies could benefit from evaluating individual sociodemographic characteristics or other metrics at finer spatial resolutions. In addition, our assessment does not account for the intricate economic and physical interdependencies across geographic regions. The power system is a complex network with interdependent relationships between transmission and distribution grids of various sizes. Although our study's original data is at the county level and considers grid connectivity as a distinguishing factor from geographic structure, future studies should incorporate more data on the transmission and distribution of power grids across regions to effectively differentiate spatial differences in power outage vulnerability.

## Methods

### Power outage data

The source of power outage information is the Environment for Analysis of Geo-Located Energy Information (EAGLE-I™) dataset. EAGLE-I™ is a geographic information system and data visualization platform developed by Oak Ridge National Laboratory (ORNL) to map populations experiencing electricity outages every 15 minutes at the county level in the United States (U.S. Department of Energy). The dataset is compiled using a variety of web parsing techniques to systematically collect near real-time outage information from several hundred large electric utilities and utility conglomerates across the U.S., which report outages from their respective collections of electric utilities (Brelsford et al., 2024). The EAGLE-I™ outage data represents approximately 90% of utility customers nationally (Brelsford et al., 2024).

The core dataset used in this study includes ten years of validated historic EAGLE-I™ records, encompassing county-level power outage information from November 2014 through December 2023 at 15-minute intervals, along with the total county-level estimated number of electricity customers. This dataset covers 3,022 counties in the contiguous United States (CONUS), representing 96.15% of the CONUS population. The data from 2014 to 2022 is publicly available at https://doi.org/10.1038/s41597-024-03095-5, with 2023 data accessible upon request from ORNL.

The data reports the estimated number of customers experiencing an outage at 15-minute intervals for all the 3,022 CONUS counties, but the total number of customers by county is not always provided. To ensure a geographically consistent measure of outage severity, we generated estimates of the total number of customers for each county. We first performed linear extrapolation to estimate state-level customer numbers from 2014 through 2016 using the state-level customer data from 2017 to 2021 provided by ORNL. Then, we calculated the average share of counties regarding the number of customers in a state using the available 2022 and 2023 county-level customer data. Finally, we used these shares to estimate county-level customer numbers for 2014 through 2021.

It is important to note that the customers in our study cannot be directly translated to the population. Electricity utilities define "customers" in various ways, typically referring to an electric meter, a building, or a facility. In residential areas, a customer might be a household, while in commercial areas, a customer could be a business or a facility.

**Power outage metrics**

To identify a county as experiencing an outage, we first calculated the power outage rate every 15 minutes. The power outage rate is defined as the ratio of the number of customers without power to the total number of customers in a county (Equation 1). This metric accounts for the total number of customers in a county and enables us to compare outage counts across counties with varying customer sizes.

$$\text{Power Outage Rate} = \frac{\text{Number of customers without power}}{\text{total number of customers}} \tag{1}$$

where the number of customers without power refers to the number of customers experiencing an outage at 15-minutes intervals, and the total number of customers represents the estimated total number of customers in the county.

We then defined a power outage event as continuously occurring whenever the power outage rate met or exceeded 0.1% based on the previous studies (Do et al., 2023; Dominianni et al., 2018). This threshold helps distinguish true power outages from other issues, such as service disconnections due to non-payment (Carr & Thomson, 2022). We screened out the 15-minute power outage records where the power outage rate was smaller than 0.1% and greater than 100% and identified a total of 3,022,915 power outage events across 3,022 counties over the 10-year period.

To better characterize these power outage events, we developed a power outage vulnerability assessment model based on the environmental exposure model (Casey et al., 2024; Miao et al., 2024; Nieuwenhuijsen, 2015), including the dimensions of frequency, duration, and intensity. We developed multiple metrics for each dimension (Fig. 1). These metrics provide a comprehensive framework for assessing power outage vulnerabilities and quantifying their impacts over time and across different regions.

- **Number of events**: this metric counts the total number of power outage events per year at the county level, measuring both the intensity and frequency of power outages.

- **Average outage rate**: this metric calculates the average power outage rate for all power outage events per year at the county level, focusing on the intensity of power outages.

- **Average/total duration**: this metric computes both the average and total duration of all power outage events per year at the county level, assessing the duration of power outages.

- **Average inter-event time**: this metric measures the average time interval between power outage events per year at the county level, assessing the frequency of power outages.

- **Number of customers affected**: this metric counts the cumulative number of customers who experienced power outages in each county per year. To provide a more complete picture, we also calculated the peak number of customers affected across all the power outage events per year at the county level. These metrics assess the intensity of power outages.

- **Cumulative user outage time**: this metric is calculated by multiplying the number of customers without power by the corresponding outage duration in a county across all power outage records per year. It measures both intensity and duration of power outages.

**Socio-economic data**

**Urban and rural classification**: urban and rural classification is commonly used to assess socioeconomic development differences at the U.S. county level (Ma & Mostafavi, 2024; Molitor, Mullins, & White, 2023). In our study, counties are categorized as either "urban" or "rural" according to the 2013 National Center for Health Statistics (NCHS) Urban-Rural Classification Scheme (National Center for Health Statistics, 2013). NCHS has developed a six-level urban-rural classification scheme for U.S. counties and county-equivalent entities. We designate counties as urban (981 counties) if they fall into the three most urban categories: large central metropolitan, large fringe metropolitan, and medium metropolitan. Counties are classified as rural (2050 counties) if they fall into the three least urban categories: small metropolitan, micropolitan, and noncore.

**Social vulnerability index**: disadvantaged communities require additional support for power outages (Bhattacharyya & Hastak, 2023; Ganz, Duan, & Ji, 2023). To identify socially vulnerable counties, we used the Social Vulnerability Index from the US Centers for Disease Control and the Agency for Toxic Substances and Disease Registry (Flanagan, Gregory, Hallisey, Heitgerd, & Lewis, 2011). The SVI aims to pinpoint areas that may require extra support during disasters. It is based on 16 census variables from the American Community Survey, including 150% below the U.S. Department of Health and Human services poverty guidelines, unemployed, housing cost burden, no high school diploma, no health insurance, aged 65 and older, aged 17 and younger, civilian with a disability, single-parent households, English language proficiency, racial and ethnic minority status, multi-unit structures, mobile homes, crowding, no vehicle, and group quarters. The SVI ranges from 0 to 1, where values closer to 0 represent lower vulnerability and values closer to 1 indicate higher vulnerability.

**Power grid connectivity data**

The level of power infrastructure vulnerability varies across different areas. This variability can be influenced by factors such as the layout of the power grid. Generally, the power grid connectivity exhibits its own distinct spatial patterns. To investigate whether different grid connectivity scenarios impact the vulnerability characteristics of power infrastructure, we collected data on two major types of grid connectivity: interconnections and the North American Electric Reliability Corporation (NERC). It is important to note that these two types of grid connectivity do not completely align with the geographic distribution of counties. Therefore, counties located in multiple grid connections are grouped together labeled as "boundary", to analyze potential differences in characteristics.

**Interconnections**: the interconnection power grids include high-voltage transmission lines, substations, and power plants spanning the United States, Canada, and parts of Mexico(United States Environmental Protection Agency). The North American grid is divided into three major interconnections: (1) Eastern Interconnection covers the eastern United States and parts of Canada (2,363 counties in our study); (2) Western Interconnection includes the western United States, parts of Canada, and some areas of Mexico (377 counties); (3) Texas Interconnection, commonly known as the Electric Reliability Council of Texas (ERCOT), covers most of Texas and operates largely independently from other interconnections (180 counties). These interconnections work together to balance supply and demand, manage power flow, and enhance grid reliability across North America.

**NERC**: NERC is a nonprofit organization created by the electric utility industry to ensure the reliability and adequacy of bulk power transmission systems in North America (North American Electric Reliability Corporation). NERC covers the interconnected power systems of the contiguous United States, Canada, and part of Mexico's Baja California. NERC's responsibilities include developing operational standards, monitoring and enforcing compliance, assessing resource adequacy, and providing education and training resources to maintain the proficiency of power system operators. The seven major NERC regional reliability entities in the U.S. are: Midwest Reliability Organization (MRO, 360 counties), Northeast Power Coordinating Council (NPCC, 117 counties), Reliability First Corporation (RFC, 416 counties), SERC Reliability Corporation (SERC, 870 counties), Southwest Power Pool (SPP, 138 counties), Texas Reliability Entity (TRE, 180 counties), and Western Electricity Coordinating Council (WECC, 377 counties).

**Statistical analysis**

The power law distribution is often used to model scenarios where a small number of events are common, while a large number of events are rare but significant (Clauset, Shalizi, & Newman, 2009). Previous research indicates power law behaviors are prevalent in blackout events (Carreras et al., 2016; Chen et al., 2001). Consequently, we applied the power law model to analyze whether it could shed light on the distribution of time intervals between successive power outage events. The power law distribution in our study is defined by Equation (2):

$$P(x) \sim x^{-\alpha} \tag{2}$$

where $P(x)$ is the probability of observing an inter-event time $x$, and $\alpha$ is the exponent of the power law. The probability decreases as $x$ increases, with the rate of decrease determined by $\alpha$. We divided county-level average inter-event time into two groups (2014-2018 and 2019-2023), and then checked the fitness of power law distribution for each group (Fig. 5e).

We employed an ordinary least squares (OLS) regression model (Equation 3) to capture the relationships between the social vulnerability index and power outage metrics (outage duration, outage rate, inter-event time, and number of events) at the county level for the two-years groups (Fig. 6).

$$y_i \sim \beta_0 + \beta_1 x_i + \varepsilon_i \tag{3}$$

where, $y_i$ is the power outage metrics of county $i$; $x_i$ is the social vulnerability index of county $i$; $\beta$ are coefficients; and $\varepsilon_i$ is the error term.

We also utilized several statistical tests to assess the significance. First, we used the paired Wilcoxon test, a non-parametric method for comparing paired observations, to analyze the box plots of the average outage rates for the 2014-2018 and 2019-2023 groups (Fig. 4a). Second, we used the Kolmogorov-Smirnov test to compare the empirical distribution with the fitted power law model (Fig. 5e), where a smaller KS statistic indicates a better fit. Third, the Kruskal-Wallis H test was employed to determine the significance of differences among the five two-year groups of power outage metrics shown in Fig. 8. This test assesses whether there are statistically significant differences among three or more independent groups. Finally, we applied the Mann-Whitney U test to evaluate differences between the distributions of two independent groups: the 2014-2018 and 2019-2023 duration proportion groups (Fig. 3b) and the urban and rural groups (Fig. 9). All analyses were conducted using Python, and the 2019 TIGER/Line US County Shapefiles were utilized to create the nationwide maps for this study (U.S. Census Bureau, 2019).

**Data availability**

The datasets used in this paper are publicly accessible and cited in this paper.

**Code availability**

The code that supports the findings of this study is available from the corresponding author upon request.

**Acknowledgments**

This work was supported by the National Science Foundation under Grant CMMI-1846069 (CAREER). Any opinions, findings, conclusions, or recommendations expressed in this research are those of the authors and do not necessarily reflect the view of the funding agencies.

**Competing interests**

The authors declare no competing interests.

**Additional information**

Supplementary material associated with this article can be found in the attached document.